\documentclass[12pt]{article}
\begin{document}
\title{Oscillation frequency of $B$ and $\overline{B}$ mesons in a QCD potential model with relativistic effect.}
\author{ $^{1}$Krishna Kingkar Pathak and $^{2}$ D K Choudhury \\
$^{1}$Deptt. of Physics,Arya Vidyapeeth College,Guwahati-781016,India\\
e-mail:kkingkar@gmail.com\\
$^{2}$Deptt. of Physics, Gauhati University, Guwahati-781014,India}
\date{}
\maketitle
\begin{abstract}

  Wavefunction at the origin with the incorporation of relativistic effect leads to singularity in a specific potential model. To regularise the wavefunction, we introduce a short distance scale here and use it to estimate masses and decay constants of $B_{d}$ and $B_{s}$ mesons within the QCD potential model.These values are then used to compute the oscillation frequency $\Delta m_{B}$ of $B_{d}$ and $ B_{s}$ mesons. The values are found to be in good agreement with experiment and other theoretical values.

\end{abstract}

The investigation of weak decays of mesons composed of a heavy quark and antiquark
gives a very important insight in the heavy quark dynamics. The study of mixing and decay constants of the $B$ meson, provides us useful information about the dynamics of quark and gluons at the hadronic scale. The weak eigenstates of neutral mesons are different to their mass eigenstates. This leads to the phenomenon of mixing whereby neutral mesons oscillate between their matter and antimatter state. This was observed first in the Kaon sector and subse-quently in $B_{d}$ and $B_{s}$ mesons. The mass difference $\Delta m_{B}$ is a measure of the frequency of the change from a $B$ into a $\overline{B}$ and so called the oscillation frequency. The decay constants of heavy mesons, one of the input parameters for oscillation frequency are crucial for interpreting data on particle-antiparticle mixing in  the neutral B meson system, and for anticipating and interpreting new signatures for CP violation. \\ 
If the CKM element is well known from other measurements, then the pseudoscalar decay constant $ f_{p}$ can be well measured. If, on the otherhand, the CKM element is less well or poorly measured, having theoretical input on $ f_{P}$  can allow a determination of the CKM element [1]. A measurement of the decay constant $f_{B}$ is difficult, since $B^{+}\rightarrow{l^{+}\overline{\nu_{l}}}$ is cabibo-suppressed in the Standard Model. Hence $f_{B_{q}}$ has to be provided from theory.\\
Here in this paper we calculate pseudoscalar masses $M_{B_{q}}$ and pseudoscalar decay constants $f_{B_{q}}$ to compute the oscillation frequency $\Delta m_{B_{q}}$,$q=d,s$ within the frame work of a potential model[2,3,4]. To incorporate relativistic effect, the necessity of a short distance scale, in analogy to QED is also pointed out here.\\

For the light heavy flavour bound system of $q\overline{Q}$ or $\overline{q}Q$ the hamiltonian can be written as

\begin{equation}
H=-\frac{\nabla^{2}}{2\mu}+V\left(r\right)
\end{equation}
$V\left(r\right)$ is the spin-independent quark-antiquark potential.
\begin{equation}
V\left(r\right)= V_{coul}\left(r\right)+V_{conf}\left(r\right)
\end{equation}
where $V_{coul}$ represents coulombic part of the potential and $V_{conf}$ represents the confining potential.
The vector and scalar confining potentials in the non relativistic limit reduce to[5,6]
\begin{equation}
V^{v}_{conf}(r)=\left(1-\epsilon_{1} \right)\left(br+C\right)
\end{equation}
\begin{equation}
V^{s}_{conf}(r)= \epsilon_{1} \left(br+C\right)
\end{equation}
reproducing
\begin{equation}
V_{conf}(r)=V^{s}_{conf}(r)+V^{v}_{conf}(r)=br+C
\end{equation}
and
\begin{equation}
V_{coul}(r)=-\alpha_{c}/r
\end{equation}
 where $ \alpha_{c}= \frac{4}{3}\alpha_{s}$, $ \alpha_{s}$ being the strong running coupling constant,$\epsilon_{1}$ is the mixing co-efficient and $b$, C are the potential parameter as is used in our previous work[3,4].\\

Considering the linear part of the potential as perturbation, coulombic part as parent and then by using dalgarno method, the wavefunction in the ground state is obtained as[2,3,4 ] 
\begin{equation}
\psi_{rel+conf}\left(r\right)=\frac{N^{\prime}}{\sqrt{\pi a_{0}^{3}}} e^{\frac{-r}{a_{0}}}\left( C^{\prime}-\frac{\mu b a_{0} r^{2}}{2}\right)\left(\frac{r}{a_{0}}\right)^{-\epsilon}
\end{equation}
\begin{equation}
N^{\prime}=\frac{2^{\frac{1}{2}}}{\sqrt{\left(2^{2\epsilon} \Gamma\left(3-2\epsilon\right) C^{\prime 2}-\frac{1}{4}\mu b a_{0}^{3}\Gamma\left(5-2\epsilon\right)C^{\prime}+\frac{1}{64}\mu^{2} b^{2} a_{0}^{6}\Gamma\left(7-2\epsilon\right)\right)}}
\end{equation}
\begin{equation}
C^{\prime}=1+cA_{0}\sqrt{\pi a_{0}^{3}}
\end{equation}
\begin{equation}
\mu=\frac{m_{i}m_{j}}{m_{i}+m_{j}}
\end{equation}
\begin{equation}
a_{0}=\left(\frac{4}{3}\mu \alpha_{s}\right)^{-1}
\end{equation}
\begin{equation}
\epsilon=1-\sqrt{1-\left(\frac{4}{3}\alpha_{s}\right)^{2}}
\end{equation}

Here $A_{0}$ is the undetermined factor appearing in the series solution of the Schr\"odinger equation, $\mu$ is the reduced mass and $m_{i},m_{j}$ are the costituent quarks masses .The term $\left(\frac{r}{a_{0}}\right)^{-\epsilon}$ is due to the relativistic effects.The strong running coupling constant appeared in the potential V(r) in turn is related to the quark mass parameter as[5] 
\begin{equation}
\alpha_{s}\left(\mu^{2}\right)=\frac{4\pi}{\left(11-\frac{2n_{f}}{3}\right)ln\left(\frac{\mu^{2}+M^{2}_{B}}{\Lambda^{2}}\right)}
\end{equation}
where, $n_{f}$ is the number of flavours, $\mu$ is renormalistion scale related to the constituent quark masses as $\mu=2\frac{m_{i}m_{j}}{m_{i}+m_{j}}$ and $\Lambda$ is the QCD scale which is taken as 0.200 GeV here. $M_{B}$ is the background mass related to the confinement term of the potential as $M_{B}=2.24 \times b^{1/2}=0.95 GeV$. In this calculation we have taken  $n_{f}=3$ as in ref.[5,7]and the input mass parameters as $m_{d}=0.36 GeV$, $m_{s}=0.46 GeV$ and $m_{b}=4.95 GeV$. With these values we calculate $\alpha_{s}=0.40$ for $B_{d}$ mesons and $\alpha_{s}=0.37$ for $B_{s}$ mesons. However we have also computed the coresponding results for $n_{f}=4$ also and listed in the tables.\\

One of the problems faced in the model to study masses and decay constants is the incorporation of relativistic effect in the wave function at the origin, since a singularity develops at $r=0$ $ (eq.7)$. However singularities at $r=0$, in relativistic and nonrelativistic approach of quark model [5,6,8] is not new and different regularisation of singularities has been discussed.\\
Here, in this work we use another way to regularise the wave function at the origin which have the quantum mechanical origin in QED. It is well known that relativistic wave function of the hydrogen atom too has such singularities. However such effect is noticeable only for a tiny region\cite{9},\\
\begin{equation}
2mz\alpha r\leq e^{-\left(\frac{1}{1-\gamma}\right)}\leq e^{-\frac{2}{z^{2}\alpha^{2}}}\sim 10^{-\frac{16300}{z^{2}}}
\end{equation}
Where z is the atomic number, m is the reduced mass of the hydrogen atom,$\alpha$ is the electromagnetic coupling constant and $\gamma=\sqrt{1-z^{2}\alpha^{2}}$. Using such hydrogen like properties in QCD, m, $\alpha$ and $1-\gamma$ are to be replaced by $\mu$, $\frac{4}{3} \alpha_{s}$ and $\epsilon$ respectively. Here $\alpha_{s}$ is strong coupling constants,$\epsilon=1-\sqrt{1-\left(\frac{4}{3}\alpha\right)^{2}}$ and  $\left(mz\alpha r\right)^{\gamma-1}$ changes to $\left(\frac{r}{a_{0}}\right)^{-\epsilon}$, leading to a cut off parameter $r_{0}$ upto which the model can be extrapolated $\left(r\geq r_{0}\right)$.\\
In analogy to the QED calculation $ \left(eq.14\right)$, using the typical length scale for the relativistic term $\left(\frac{r}{a_{0}}\right)^{-\epsilon}\leq \frac{1}{e} $ , we get the cut off parameter\\
\begin{equation}
r_{0}\sim a_{0}e^{-\frac{1}{\epsilon}}.
\end{equation}
Unlike QED, it is flavour dependent depending on the flavours of the quark masses$(eq.11)$. \\

The decay constants of mesons are important parameters in the study of leptonic or non-leptonic weak decay processes and in the neutral$B-\overline{B}$ mixing process. In the nonrelativistic limit, the decay constant can be expressed through the ground state wavefunction at the origin $\psi_{p}(0)$ by the Van-Royen-Weisskopf-formula [10].Though most of the models predict the mesonic mass spectrum succesfuly, there are disagreements in the predictions of there decay constants. So we reexamine the predictions of the decay constants with a new short distance scale.\\
We consider the nonrelativistic expression for $f_{p}$ as [10,11]

\begin{equation}
f_{p}=\sqrt{ \frac{12}{M_{p}}|\psi\left(0\right)|^{2}}
\end{equation}

where $M_{p}$ is the pseudoscalar mass of mesons and is calculated by using the relation[12] 
\begin{equation} 
M_{p}=m_{i}+m_{j}-\frac{ 8\pi\alpha_{s}}{3m_{i}m_{j}} |\psi\left(0\right)|^{2}
\end{equation}
\begin{table}[h]
\begin{center}
\caption{Values of $r_{0}$ and $M_{p}$ for $B_{d}$ and $B_{s}$ mesons }
\vspace {.2in}
\begin{tabular}{c c c c}\hline
mesons&$|\psi\left(0\right)|$ in $GeV^{3/2}$ & {values of $r_{0}$ in $ Gev^{-1}$}& $M_{B}$ in GeV\\\hline 

$B_{d}$&0.141 &0.009 &5.273[our work with $n_{f}=3$] \\
 &0.162&0.021&5.256[our work with $n_{f}=4$]\\
 & & &5.279 [13]\\ 
 & & &5.285 [14]\\
 & & &5.279 [15]\\\hline
$B_{s}$&0.178 & 0.002 &5.370[our work with $n_{f}=3$] \\
 &0.207&0.007&5.349 [our work with $n_{f}=4$]\\
 & & & 5.369[13]\\
 & & & 5.373[14]\\
 & & & 5.375[15]\\\hline

 \end{tabular}
\end{center}
\end{table}
The values of $r_{0}$ and coresponding wavefunction at the origin for $B_{d}$ and $B_{s}$ are clculated to study pseudoscalar masses and are shown in table 1. Using eq.16, we compute the decay constants for $B_{d}$ and $B_{s}$ mesons and put in table 2. The reults are found to be well with the experimental values.\\

 The neutral $B_{d} \rm{~and~} B_{s}$ mesons can mix with their antiparticles by means of box diagram involving exchange of a pair of W bosons and intermediate $u,c,t$ quarks leading to oscillations between the mass eigenstates. This 
mass oscillation is parametrized as the oscillation frequency or  mixing mass parameter($\Delta m$) and is given by [16,17]

\begin{equation}
\Delta m_{B}= \frac{ G^{2}_{F}m^{2}_{t}M_{B_{q}}f^{2}_{B_{q}}}{8\pi}g\left(x_{t}\right)\eta_{t}|V^{*}_{tq}V_{tb}|^{2}B,   q=d,s
\end{equation}
where $\eta_{t}$ is the gluonic correction to the oscillation and is taken 0.55 as in ref.[16], the last factor B is the bag parameter which represents the correction to the vacuum insertion and is taken as 1.34 [16]. The  function $g\left(x_{t}\right)$  is
  given by[18]
\begin{equation}
g\left(x_{t}\right)=\frac{1}{4}+\frac{9}{4\left(1-x_{t}\right)}-\frac{3}{2\left(1-x_{t}\right)^{2}}-\frac{3 x^{2}_{t}}{2\left(1-x_{t}\right)^{3}}
\end{equation}
Here, $x_{t}=\frac{m^{2}_{t}}{M^{2}_{W}}$. The values $m_{t}\left(174 GeV\right)$, $m_{W}\left(80.403 GeV\right)$,and the CKM matrix elements  $|V_{tb}|\left(1\right)$, $|V_{td}|\left(7.4\times10^{-3}\right)$,$|V_{ts}|\left(40.6\times10^{-3}\right)$ are taken from the particle data group [13].\\
 We use the estimsted values of masses and decay constants to compute the mixing mass parameters and put in the table 2. The decay constants are found to be comparable with the other theoretical values and mixing mass parameters are found to be in good agreement with the experimental values.
\begin{table}[h]
\begin{center}
\caption{Decay constant and oscillation freq. of B mesons}
\vspace {.2in}
\begin{tabular}{c c c}\hline
mesons&$f_{B}$(in GeV) & $\Delta m_{B}$(in $ps^{-1}$) \\\hline 

$B_{d}$&0.213[our work with $n_{f}=3$] &0.55[our work with $n_{f}=3$]  \\
 &0.246[our work with $n_{f}=4$]&0.74[our work with $n_{f}=4$]\\
 &0.189 [22]&0.50[13] \\
 &0.196(29)[23]&0.547[20] \\
 &$0.190(7)_{-1}^{+24}$[22]&0.515[21] \\
 &0.210(9)[25]& \\
 &0.216(9)(19)(6)[26,27]&\\\hline
$B_{s}$&0.265[ our work with $n_{f}=3$] &17.34[our work with $n_{f}=3$]  \\
 &0.311[our work with $n_{f}=4$]&23.88[our work with $n_{f}=4$]\\
 &0.218 [22]&17[13 ] \\
 &0.216[23]&17.77[19] \\
 &$0.217(6)_{-28}^{+32}$[24]& \\
 & 0.244(21)[25]& \\
 & 0.259(32)[26,27]&\\\hline
\end{tabular}
\end{center}
\end{table}

To conclude, in the present work, we have incorporated the relativistic effects to the wave function at the origin in a QCD potential model.We have introduced a short distance scale $r_{0}$ in analogy to QED.
Unlike QED, in QCD such short distance scale become flavour dependent. As expected, its magnitude is far larger than its QED counterpart but still far smaller than the measure of finite size of hadrons or its constituents. It is however well within the reach of LHC[26] where distance down to a scale as short as  $5\times 10^{-20} m $ -$10^{-21} m $ will be explored. Theoretically this short distance scale $r_{0}$ can be roughly associated with the ultraviolate regularisation scale of QCD.\\

We have found from the calculation that generally the masses are not so sensitive to running coupling constants since the heavy quark constituent masses dominent here, but the decay  constants are very much  sensitive to the running coupling constants and therefore a slight change in $\alpha_{s}$ as well as $r_{0}$ deviates the results significantly. \\
In table 2 we confront our results for pseodoscalar decay constants and oscillation frequency($\Delta m_{B}$) of the $B_{d}$ and $B_{s}$ mesons with the recent predictions of quark model[22], three flavour lattice QCD [27,28], Collaboration[19,20,21] and available experimental data[13]. In our calculation we have found  $\frac{f_{B_{s}}}{f_{B}}$=1.24  which is in accordance with the lattice result $\frac{f_{B_{s}}}{f_{B}}$=1.20(3)(1) [26,27].  
The results in table 1 and table 2 for $B_{d}$ and $B_{s}$ mesons with the short distance scale is found to be in good agreement between all presented theoretical predictions as well as experimental data for $n_{f}=3$ and the corresponding results with $n_{f}=4$ are found to overshoot the theoretical predictions and experimental data. \\

 \end{document}